# Improved Method of Stock Trading under Reinforcement Learning—Based on DRQN and Sentiment Indicators ARBR

**Peng Zhou[1a], Jingling Tang[1b]**

[1]Chengdu University of Technology, Chengdu, China

[a]zhou.peng@student.zy.cdut.edu.cn, [b]tjlcdut@163.com

Abstract—With the application of artificial intelligence in the financial field, quantitative trading is considered to be profitable. Based on this, this paper proposes an improved deep recurrent DRQN-ARBR model because the existing quantitative trading model ignores the impact of irrational investor behavior on the market, making the application effect poor in an environment where the stock market in China is non-efficiency. By changing the fully connected layer in the original model to the LSTM layer and using the emotion indicator ARBR to construct a trading strategy, this model solves the problems of the traditional DQN model with limited memory for empirical data storage and the impact of observable Markov properties on performance. At the same time, this paper also improved the shortcomings of the original model with fewer stock states and chose more technical indicators as the input values of the model. The experimental results show that the DRQN-ARBR algorithm proposed in this paper can significantly improve the performance of reinforcement learning in stock trading.

## 1. Introduction

In recent years, with the widespread application of artificial intelligence technology, various quantitative trading methods represented by reinforcement learning have also become a hot topic in the financial field [1]. Reinforcement learning is one of the excellent methodologies born with the development of artificial intelligence, and it has been used maturely in the areas of computer vision and game science [2]. To adapt to the complex financial environment, researchers have tried to use the mechanism of reinforcement learning to select different actions from the economic climate, and through learning to find an optimal strategy that adapts to the current environment, so as to explore the application of deep reinforcement learning to the financial market—possibility to provide counseling ideas for individuals and institutions to allocate and trade financial assets [3]. Bekiros (2010) proposed an intelligent trading system based on reinforcement learning and supplemented by an adaptive network-based fuzzy inference system for high-frequency trading [4]; Liu Quan et al. (2017) evaluated the performance of different strategic games under DRQN training and pointed out that DRQN is better than DQN model in terms of the learning curve and reward situation [5]; Xiang Gao (2018) compares the benefits of DRQN model under different architectures with two methods of univariate and bivariate, and concludes that the LSTM architecture in both cases have good results [6]; Uriel Corona-Bermudez et al. (2020) used deep Q and GRU network to build an asset management model based on cryptocurrency and achieved good returns [7]; Liu Wei (2021), according to the results of sentiment analysis of stock market data and comment data, designed a stock trading strategy model based on the deep Q network to make the stock trading model more stable [8]; The paper from Dai Xiaoxue (2021) proves the performance of reinforcement learning methods in selecting stocks is better than the buy and hold strategy and MACD strategy [9].

It can be seen from the above literature that the deep reinforcement learning method has made specific achievements in the financial field, especially in the aspect of trading strategies. Nevertheless, this research also found that most stock strategy models blindly pursue the mathematical laws of stock price changes and ignore the impact of irrational investor behavior on the market, resulting in unsatisfactory application effects in a non-efficiency stock market environments like China. However, it is still doubtful whether a single model can be applied to the complex financial environment. Based on this, this article uses the DRQN recurrent network and the sentiment indicator ARBR to construct a trading model for the stock market. At the same time, this paper improves the shortcomings of the traditional trading model with fewer stock states, chooses more technical indicators as the input value of the model, and uses DRQN to solve the limited memory of empirical data storage and the observable Markov property of the traditional DQN stock trading model. The experimental results show that the DRQN-ARBR model can significantly improve the performance of reinforcement learning in stock trading.

## 2. Algorithm principle

### 2.1. Q-Learning algorithm

The learning process of Q-learning algorithm can be described as follows: In the target environment, the agent can take action space $A$ and state space $S$. The probability transition matrix $P$ can make the transition from the current state to the next state and obtain a reward $R$. The Q value from the corresponding state to the action is represented by $Q(s,a)$.

Suppose the state of the Agent at time $t$ is $s_t$, and the action taken at this time is $a_t$. After the agent executes the action, it shifts to time $t+1$, the state changes to $s_{t+1}$, and the reward it gets is $r_t$. By updating the value of $Q(s,a)$ through all the recorded $(s_t, a_t, s_{t+1}, a_{t+1})$, iteratively find the optimal strategy. The formula can be expressed as:

$$\widehat{Q}(s,a) = \widehat{Q}(s,a) + \alpha\left(r + \gamma\max_a \widehat{Q}(s',a) - \widehat{Q}(s,a)\right) \quad (1)$$

Q-learning algorithm is a reinforcement learning algorithm based on Q value. It forms a Q table by the variables of state and action, aims at storing the Q value under the corresponding combination, and then selects the action that can get the maximum reward through the Q value [10].

### 2.2. Deep Recurrent Q network

Minih et al. combined CNNs and the Q learning algorithm in traditional RL and proposed a deep Q network DRL algorithm. This algorithm solves algorithm instability when a nonlinear function approximator is used to represent the value function to a certain extent. DQN is a type of DRL, which is a combination of deep learning and Q learning. When the combination of state and action is inexhaustible, the traditional Q-learning algorithm can no longer select the optimal action by looking up the Q table. The optimal wireless approximation can be found through the deep Q network without exhaustively enumerating all the combinations [11].

In the DQN neural network, the input is the state $s_1$ and the action space $\{a_1, a_2, ..., a_n\}$, and the output is the Q value $q(s_1, a_1), q(s_1, a_2), ..., q(s_n, a_n)$ corresponding to each action. After that, we only need to select the action with the largest Q value to operate. The formula for determining the Q value is:

$$Q(s_t, a_t) = R_{t+1} + \gamma \max_a Q(s_{t+1}, a) \quad (2)$$

Its principle lies in that in the action space, each action is predicted by neural network to get the Q estimated value, and the action with the largest Q estimated value is selected to get the corresponding reward. This relevant chart is shown in Figure 1.

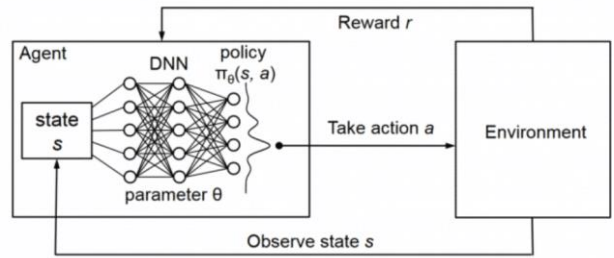

Fig.1 chart of deep Q network

## 3. Model based on deep recurrent Q network and sentiment inicator ARBR

*3.1. model structure*

As shown in Figure 2, this article designs a stock buying and selling point judgment method based on the ARBR sentiment indicator and a deep recurrent Q fusion framework to make the strategy more stable in quantitative trading in the stock market. The flow chart based on the DRQN-ARBR model is shown in Figure 3, which mainly includes three parts:

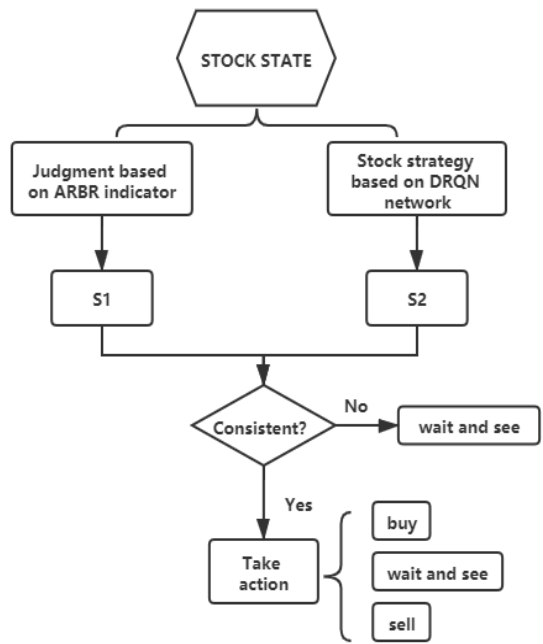

Fig.2 Model structure based on deep recurrent Q network and ARBR indicator.

(1) *Stock State*: By preprocessing the stock market data in groups and preprocessing the daily technical indicator data and ARBR values, we use all the integrated data as the spatial state of the stock.

(2) *A Judgment model of Buying and Selling Points Based on ARBR Sentiment Indicator:* According to the spatial state of stocks, we construct a dynamic stock trading model based on sentiment indicators. Through the dynamic comparison of the stock AR and BR values, the best operation at the current time is judged, and the corresponding operation signal $S1$ is issued.

(3) Stock trading model based on DRQN network: We build a trading model based on the DRQN neural network through the state of the stock and obtain the buying and selling signal $S2$ through this model. If it is consistent with the previously sent $S1$ signal, perform the corresponding action.

*3.2. stock state*

To better represent the complex financial market, this paper includes as many relevant factors affecting stock trading behavior as possible to show the state of stocks. After many experiments, this research finally decided to use group closing prices, stock technical indicators, and market sentiment as the state space of the target stocks. The specific construction diagram is shown in Figure 3:

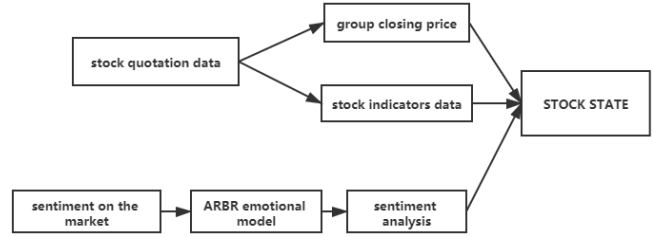

Fig.3 Construction diagram of stock state

*3.2.1. group closing price*

Since this article selects a graded high-frequency stock price data within five years as the experimental data, to process the enormous stock data, we use a grouping method for preprocessing: Taking 30 minutes as a group, the opening price of the first member of each group is used as the opening price of the group, and the closing price of the last member is the group's closing price. That is to say, the closing price of the last group is the opening price of the current group, so in the process of data storage, we only need to record the pair of data at different times and the current closing price.

At the same time, we calculate the logarithmic rate of the adjacent eight groups' closing prices and regard the result as the eight features of the stock state of the current group. The characteristics of these states will be standardized by z-score and then fed back to the LSTM network for processing. The z-score processing formula is:

$$z = \frac{x - \mu}{\sigma} \tag{3}$$

Among them, $x$ is the original data, $\mu$ is the average data, and $\sigma$ is the standard deviation.

*3.2.2. Stock indicators data.*

The stock index data is used as a technical explanation of stock price changes, which includes many professional stock technical indicators, such as MACD, SMA, MFI and RSI.

*3.2.3. Quantification of market sentiment.*

There is a significant correlation between the sharp rise or fall of stock price and the change of public mood. In this paper, we use market transaction data to indirectly reflect market sentiment changes.

The sentiment index ARBR is composed of the popularity index AR and the willingness index BR. AR and BR are a pair of data reflecting the strength of long and short forces at different times by analyzing historical stock prices. They can reasonably infer the current market sentiment and predict the reversal point of the stock price more accurately.

The AR indicator reflects the popularity of the target stock in the trading market by comparing the changes in the opening price with the highest price and the lowest price in a certain period. The formula is:

$$AR = \frac{\sum_{i=1}^{N}(\beta_i - \alpha_i)}{\sum_{i=1}^{N}(\alpha_i - \mu_i)} \times 100 \tag{4}$$

The BR indicator reflects the degree of willingness to buy and sell the target stock by comparing the position of the closing price in a certain period in the price fluctuation of the period. The formula is:

$$BR = \frac{\sum_{i=1}^{N}(\beta_i - \alpha_i')}{\sum_{i=1}^{N}(\alpha_i' - \mu_i)} \times 100 \tag{5}$$

Where $\alpha_i$ is the opening price on the i-th day, $\beta_i$ is the highest price on the i-th day, $\mu_i$ is the lowest price on the i-th day, $\alpha_i'$ is the closing price of the day before the i-th day, and $N$ is the given time period.

*3.3. the action state of the stock.*
The Agent can choose to execute at each time point: buy, wait and see, and sell, which are represented by numbers 1, 0, and -1 respectively. The action state can be expressed as:

$$a = \begin{cases} 1 \\ 0 \\ -1 \end{cases} \tag{6}$$

It is worth noting that the number 1 represents the purchase of stocks with a specified transaction amount at the current moment and the payment of transaction fees; the number 0 represents wait-and-see, that is, neither buy nor sell; the number -1 represents sell stocks with a specified transaction amount at the current time and pay transaction fees. In this study, we set the cost of each transaction as 0.1% of the transaction amount. In addition, the model assumes that it operates in an ideal environment, which means that the transaction price of the order is the actual transaction price. However, in the actual stock market this is not necessarily the case.

*3.4. the reward function*
The reward value refers to the reward that the agent gets after executing each action. The model optimizes the accuracy of the model and adjusts the subsequent actions by setting the reward value. Because this model studies minute-level high-frequency stock data, we define the maximum profit per minute after grouping as the reward value of the model.

$$r_t = P_t - P_{t-1} \tag{7}$$

Among them, $r_t$ is the reward value currently obtained, $P_t$ is the price at time $t$, and $P_{t-1}$ is the price at the previous time. The reward value is the difference between the prices at the two moments. When the current price is greater than the past price, the reward value is positive; when the current price is lower than the past price, the reward value is negative. Since the strategy of the model is to find the return under the best strategy, the final cumulative return is:

$$R_t = \sum_{t=1}^{T} r_t \tag{8}$$

## 4. Example analysis

*4.1. Data selection and processing.*
The experimental data in this paper comes from China Securities Net and Tushare big data community. We select one-minute high-frequency stock data of Tapai Group (002233) from January 2017 to July 2021. We use a grouping method to preprocess the vast data. Take 30 minutes as a group, and use the first data and the last data of each group as the opening price and closing price, respectively. Therefore, the empirical process only needs to record the closing price every 30 minutes. At the same time, we record the logarithmic return rate of each group and use the z-score method to standardize the data as the state of each group. In addition to using the basic stock prices as market data, this article also integrates 20 technical indicators as one of the input states of the model. Figure 4 is a sequence diagram of selected stock price changes over the past five years.

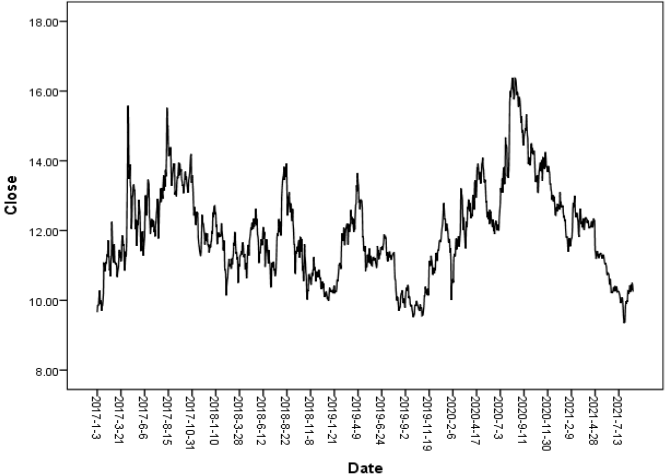

Fig.4 Five-year time chart of tapai group

### 4.2. model training

According to the sequence diagram, we record the AR and BR values of the target stock at different times, as shown in Figure 5:

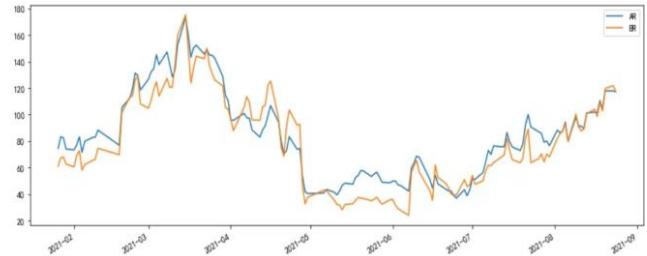

Fig.5 Ar and br values of the target stock at different times

    Using the ARBR rules set up in the paper, we train the stock trading model guided by the ARBR sentiment indicator, and visualize the trading situation as shown in Figure 6. It can be seen from the figure that the green is the stock selling point, and the red is the stock buying point. In the whole picture, the selling points of stocks are basically concentrated on the upper part of the chart, while the buying points are mostly at the bottom of the chart. At the same time, the forecast can be found in the face of huge changes in the stock price it seems to be extremely accurate, which shows that the ARBR sentiment indicator can well control the turning point of stocks and perform extremely well in the prediction of extreme sentiment reversals.

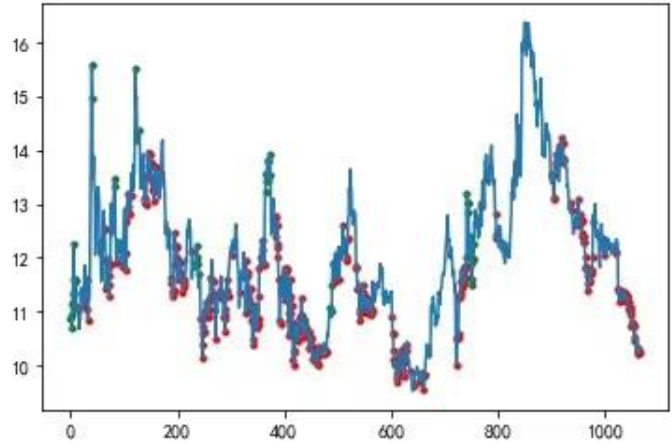

Fig.6 Visualized buying and selling under emotiona arbr indicators.

After data preprocessing and basic setting of ARBR emotional indicators, we input the data of training indicators into the DRQN-ARBR model for training. The setting values of model indicators are shown in Table 1.

Table.1 setting values of the model.

| Training index | Set value |
| --- | --- |
| batch size | 16 |
| learning rate | 0.00025 |
| gamma | 0.001 |
| hidden layer | 32 |

We use the buy-and-hold strategy as the baseline and compare it with the earnings of the paper model. The results are shown in Figure 7:

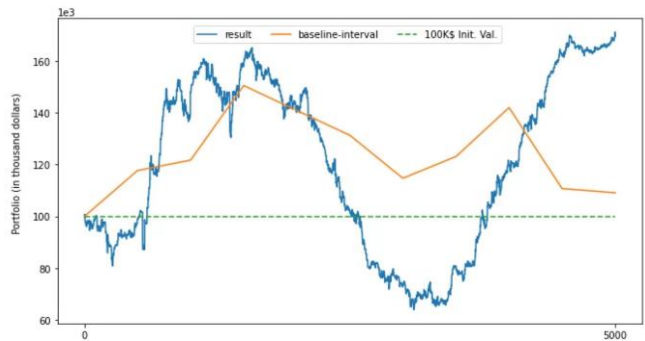

Fig.7 Compare performance

After comparing the two return curves, it is not difficult to find that the DRQN-ARBR model's overall return is greater than that of the buy-and-hold strategy, and in the end it gained more than 170,000 yuan. At the same time, we also summarized the quantitative benefits of other strategic models in the past five years, as shown in Table 2:

Table.2 different model performance evaluation.

| index | accumulated income |
| --- | --- |
| ARBR-DRQN | 174312.5 |
| DRQN | 159603.24 |
| LSTM | 73413 |
| MACD | 34200 |

It can be seen from the table that all four strategies have achieved positive returns. Among them,

the strategy based on ARBR-DRQN has achieved the best profit among all strategies, followed by DRQN, and the MACD strategy is the worst. And the DRQN-ARBR model has the best response speed and the recognition of buying and selling points compared to other models, so the algorithm proposed in this paper is effective.

## 5. Conclusion

This paper proposes an improved reinforcement learning method based on DRQN-ARBR and tests its performance with other strategy models in stock trading. After empirical research, it is concluded that the model in the paper is superior to the buy-and-hold strategy and other traditional stock quantitative trading strategies such as LSTM and MACD in terms of return. It is worth noting that the model considers the impact of irrational investors on the market and solves the problems of the traditional DQN model with limited experience storage memory and the impact of observable Markov properties on the market performance. Due to the complexity of the financial market, the next research will be used to discuss whether this model can achieve the same good results in financial trading markets other than stocks and how to adjust the model parameters to improve performance.